\documentclass[%
reprint,
superscriptaddress,
nofootinbib,
amsmath,amssymb,
aps,
floatfix
]{revtex4-2}

\usepackage[pdftex]{graphicx}
\usepackage{bm}

\usepackage[utf8]{inputenc} 

\usepackage[colorlinks=true,linkcolor=blue]{hyperref}

\usepackage{mathptmx}
\usepackage{etoolbox}
\usepackage{enumitem}

\newcommand{\un}[1]{\>\mathrm{#1}}

\begin{document}
\title{Shock-induced melting and crystallization in titanium irradiated by ultrashort laser pulse}

\author{Vasily Zhakhovsky}
\email{basi1z@ya.ru}
\affiliation{Joint Institute for High Temperatures, RAS, 13/2 Izhorskaya st., 125412 Moscow, Russia}
\affiliation{Dukhov Research Institute of Automatics, 22 Sushchevskaya st., 127030 Moscow, Russia}

\author{Yury Kolobov}
\email{yuri4877@mail.ru}
\affiliation{Federal Research Center of Problems of Chemical Physics and Medicinal Chemistry, RAS, 1 Semenova av., 142432 Chernogolovka, Russia}
\affiliation {Lomonosov Moscow State University, Leninskie Gory, 119991 Moscow, Russia}

\author{Sergey Ashitkov}
\email{ashitkov11@yandex.ru}
\affiliation{Joint Institute for High Temperatures, RAS, 13/2 Izhorskaya st., 125412 Moscow, Russia}

\author{Nail Inogamov}
\affiliation{Landau Institute for Theoretical Physics, RAS, 1A Semenova av., 142432 Chernogolovka, Russia}
\affiliation{Joint Institute for High Temperatures, RAS, 13/2 Izhorskaya st., 125412 Moscow, Russia}
\affiliation{Dukhov Research Institute of Automatics, 22 Sushchevskaya st., 127030 Moscow, Russia}

\author{Ivan Nelasov}
\affiliation{Federal Research Center of Problems of Chemical Physics and Medicinal Chemistry, RAS, 1 Semenova av., 142432 Chernogolovka, Russia}

\author{Sergey Manokhin}
\affiliation{Federal Research Center of Problems of Chemical Physics and Medicinal Chemistry, RAS, 1 Semenova av., 142432 Chernogolovka, Russia}

\author{Victor Khokhlov}
\affiliation{Landau Institute for Theoretical Physics, RAS, 1A Semenova av., 142432 Chernogolovka, Russia}

\author{Denis Ilnitsky}
\affiliation{Dukhov Research Institute of Automatics, 22 Sushchevskaya st., 127030 Moscow, Russia}

\author{Yury Petrov}
\affiliation{Landau Institute for Theoretical Physics, RAS, 1A Semenova av., 142432 Chernogolovka, Russia}

\author{Andrey Ovchinnikov}
\affiliation{Joint Institute for High Temperatures, RAS, 13/2 Izhorskaya st., 125412 Moscow, Russia}

\author{Oleg Chefonov}
\affiliation{Joint Institute for High Temperatures, RAS, 13/2 Izhorskaya st., 125412 Moscow, Russia}

\author{Dmitriy Sitnikov}
\affiliation{Joint Institute for High Temperatures, RAS, 13/2 Izhorskaya st., 125412 Moscow, Russia}

\begin{abstract}
Modification of titanium microstructure after propagation of a melting shock wave (SW) generated by a femtosecond laser pulse is investigated experimentally and analyzed using hydrodynamic and atomistic simulations.
Scanning and transmission electron microscopy with analysis of microdiffraction is used to determine the microstructure of subsurface layers of pure titanium sample before and after modification.
We found that two layers of modified titanium are formed beneath the surface. A top surface polycrystalline layer of nanoscale grains is formed from a shock-molten layer via rapid crystallization. In a deeper subsurface layer, where the shock-induced melting becomes impossible due attenuation of SW, recrystallization of plastically deformed titanium leads to grain size changes in comparison with intact titanium.
Molecular dynamics simulation of single-crystal titanium reveals that the SW front continues to melt/liquefy even after its temperature drops below the melting curve $T_m(P)$. The enormous shear stress generated in a narrow SW front leads to collapse/amorphization of the crystal lattice and formation of a supercooled metastable melt. Such melt crystallizes in an unloading tail of SW until its temperature becomes higher than $T_m(P)$ due to a rapid pressure drop. Later, crystallization of the subsurface molten layer will continue after the heat leaves it. After the shear stress drops below $\sim 12 \un{GPa}$ within the SW front, such the cold mechanical melting ceases giving place to the shock-induced plastic deformations. The depth of modification is limited by SW attenuation to the Hugoniot elastic limit, and can reach several micrometers. The obtained results reveal the basic physical mechanisms of surface hardening of metals by ultrashort laser pulses.
\end{abstract}

\keywords{
laser pulse \sep shock wave \sep shock melting \sep crystallization \sep laser shock peening
}

\date{\today}
\maketitle

%
\section{Introduction}
\label{sec:intro}

Improvement of metal tools utilized in technical and medical applications by modification of their working surfaces with ultrashort deposition of laser energy appears to have considerable promise for high-power pulsed technologies for processing of structural and functional materials.
It was noted in the review \cite{Kolobov:2018} that the nanosecond  laser pulses can be used to increase surface hardness and wear resistance, increase corrosion resistance and modify other properties of near-surface layers of materials. Here we will refer such pulses as short. By contrast, exposure to pulses with picosecond and even shorter duration (ultrashort pulses) provides much higher energy density deposited within a thin surface layer, which can trigger more intense modification there.

The processes induced by pulsed laser irradiation can be utilized to form new materials and tools with the unique mechanical, thermal and other features. By varying the parameters of laser irradiation such processes may provide the controlled modification of a phase-structural state of thin near-surface layers and also may change the surface morphology by formation of voids, craters and jets \cite{Ashitkov:2012:JETPLett,Loktionov:2014:HighTemp,Inogamov:2016:Nanoscale}. Another feasible laser-induced modification of surface shape is a multimodal roughness, which can be used to produce the superhydrophobic or superhydrophilic properties \cite{Vorobyev2013-obzor=optika+wetting}.
Moreover, the colored painting can be depicted on surface by laser pulses as it was demonstrated on titanium with using the controlled exposure to scanning laser beam consisting of pulses of nanosecond duration \cite{Veiko:21}.

It was demonstrated that the usage of higher intensity laser pulses may result in not only ablation of surface material but also in generation of high-pressure shock waves (SW) required for the laser shock peening (LSP) \cite{Fabbro:1990,Correa:2015,Tokmacheva-Kolobova:2021}. Such SW processing opens doors for improving the mechanical characteristics of both the thin surface layers (less than one micrometer thick) and near-surface layers of materials to a depth from one to several hundred micrometers \cite{Kolobov:2018,JIA2014354}.
Recently we have shown that the formation of such near-surface layers with nanocrystallines produced by exposure to laser pulses of nanosecond duration allows multiply (not less than twice) increase the multi-cycle fatigue resistance of plate-like samples of technically pure titanium (VT1-0) up to 1 mm thick \cite{kolobov2021studying,Kolobov:2022}.

Last decade the pico- and  femtosecond laser pulses were suggested for the LSP and tested on different materials in pioneer works \cite{Tsujino:2012,Sano.T_fs.LSP_APL:2014,fs-LSP-Sano-2018a,fs-LSP-Sano-2018b}.
In contrast to the nanosecond LSP, the pico- and  femtosecond laser pulses can produce much stronger shocks with pressure of several orders of magnitude higher under target surfaces without protection by films or covering water \cite{T.Sano:2017,Kudryashov:2020,Yiling:2021,Yuxin:2021}.
The optical breakdown and filamentation of laser beams in the covering materials strongly limits the light intensity reaching a target during nanosecond exposure. As a result, the generated pressures cannot exceed several gigapascals (GPa) in the nanosecond LSP. Whereas the pressures up to 1000 GPa are easily achieved by using the ultrashort pulses for LSP if the laser beam propagates through air or vacuum due to their high optical strength, for which the energy absorbed in a surface layer can reach up to $F_{abs} \sim 10^6 \un{J/m^2}$ \cite{Yiling:2021,Yuxin:2021}.

Despite many published works on the LSP, the structure features of a subsurface layer just beneath the surface irradiated by laser pulses of femtosecond duration remain virtually unstudied. Only in a few works the investigation of the extracted thin foils (lamellae) from transverse sections using the focused ion beams \cite{Ye:2014240} were carried out with using the transmission electron microscopy (TEM). But even in such works the considerable attention has not been given to the study of the microstructure of subsurface layers \cite{Zhang:2019573}.

Melting process is of considerable importance in many laser applications. As a rule in the corresponding works, melting is analyzed via simple consideration of energy deposition and heat transfer due to thermal conductivity.
In the present work, the non-equilibrium melting induced by shock loading via dissipation of kinetic energy within a SW front, where a shear stress exceeds the shear modulus of metal, and following crystallization are in the focus of our study.
Melting SWs have been studied in detail in \cite{Eliezer:1993,Povarnitsyn:2008,Budzevich:2012,Qiu:2021}, but these works have not addressed to hardening processes in materials after shock loading.
A ultrashort pulse heating of metals with the relatively low electron heat conductivity (such as titanium and zirconium) produces a thin heated surface layer and thus a thin molten layer. On the other hand, the thickness of shock-molten near-surface layer produced just within a SW front can be much thicker due to the slow attenuation of SW pressure if the initial pressure depending on the absorbed laser energy is high enough.

In this work we show that melting and subsequent crystallization of the near-surface layer radically alters the crystal structure of this layer. In particular, the ultra-high quenching rate of the molten near-surface layer in most known cases leads to refinement of its grain structure up to formation of nanocrystallines there. The latter allows, along with a significant hardening, to maintain the necessary level of plasticity, or even increase it.
This is a fundamentally important achievement for practical applications, since it makes possible a deep-layer ultrashort LSP with the record hardening in the near-surface layers of metals.

The aim of this work is to study the structural-phase transformations in the sub- and near-surface layers of technically pure recrystallized titanium (alloy VT1-0) after exposure by the single femtosecond laser pulses of terawatt energy, generated by our chrome-forsterite laser system. We performed the two-temperature hydrodynamics modeling followed by molecular dynamics simulation of titanium to demonstrate that after such fast energy deposition a shock wave with pressure of several hundreds giga pascal is generated, which is enough for melting within a shock jump. This suggests the manifestation of previously unknown features in formation of microstructure after phase transformations in the subsurface layers of the alloy under study, as well as a significant influence of material modification in those layers on the mechanical properties of bulk samples. Indeed, the deep structure modifications of surface layers are found using scanning and transmission electron microscopy with analysis of microdiffraction. In particular the top polycrystalline layer of nanoscale grains was detected. Also we found a deeper subsurface layer, where the shock-induced melting front cannot propagate, with recrystallized titanium having the crystalline grains finer than in the initial titanium.

\section{Material and experimental technique}
\label{sec:method}

\subsection{Preparation of titanium samples}
Commercially pure titanium VT1-0 ($\alpha$-titanium with HCP lattice) was chosen as the material for research, the chemical composition of which is well known \cite{Kolobov:2018}.
Samples for research were made from rods (8 mm in diameter) of submicrocrystalline (SMC) VT1-0 alloy with the size of grain-subgrain structure elements of about $0.2 \un{\mu m}$. The SMC column structure was formed by sequential actions of cross-helical and section rolling according to the method described in \cite{Kolobov:2018}. In the samples of the alloy with the initial SMC structure it is possible to obtain a volume-uniform material having a polycrystalline structure with the average grain size of $35\pm 3\un{\mu m}$ by subsequent final recrystallization annealing at the temperature of 1123 K for 1 hour.

\begin{figure}[t]
\centering\includegraphics[width=0.85\columnwidth]{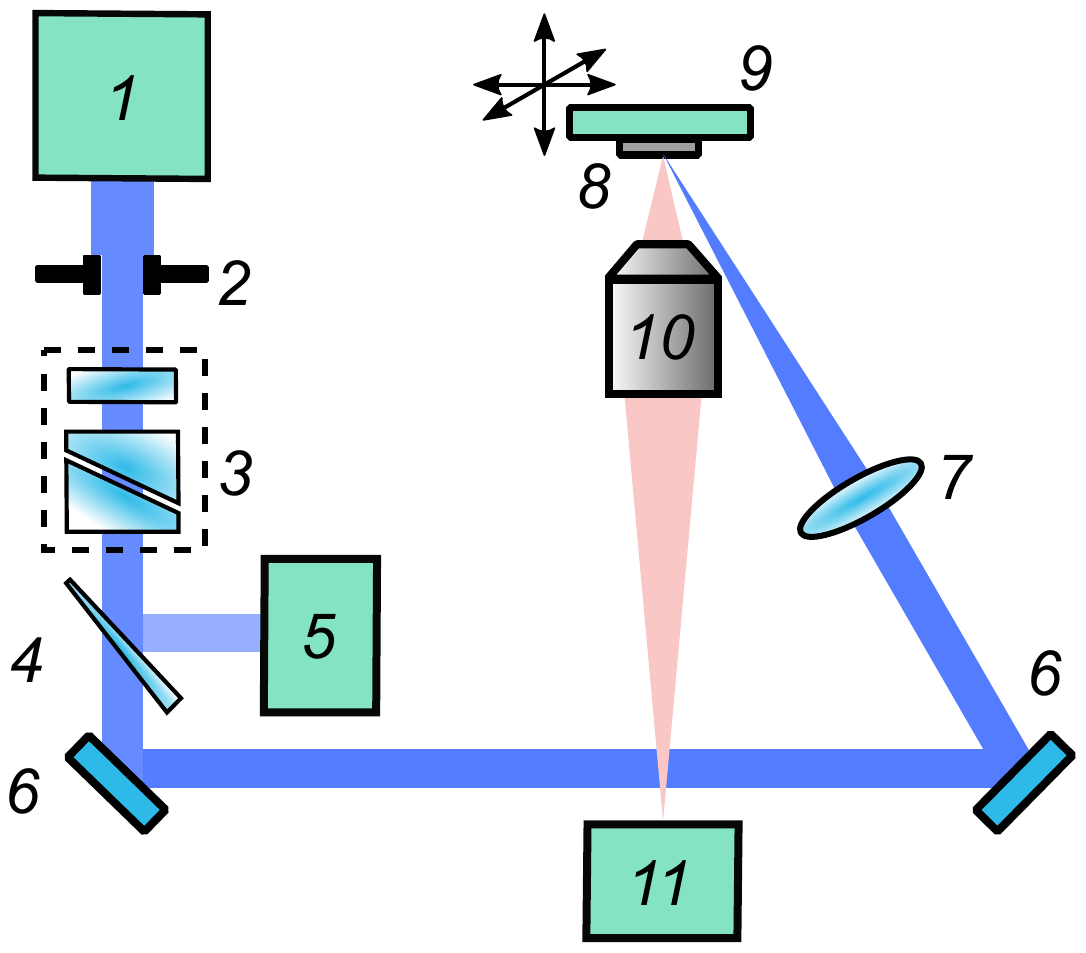}
\caption{\label{fig:01}  Experimental technique of laser processing: 1 -- laser radiation source, 2 -- aperture, 3 -- polarized attenuator, 4 -- wedge-shaped glass plate, 5 -- photo-diode, 6 -- aligner mirrors, 7 -- focusing lens, 8 -- sample, 9 -- three-axis coordinate worktable, 10 -- micro-lens, 11 -- CCD camera. }
\end{figure}

\subsection{Laser processing technique}

A terawatt femtosecond chromium-forsterite laser system shown in Fig.~\ref{fig:01}(1) was employed for laser processing of titanium samples. It generated laser pulses at a wavelength of 1240~nm with pulse duration of 110~fs (by the FWHM level) and the energy up to $30\un{mJ}$\cite{Agranat_Cr:f:QE:2004}. The sample surface was exposed to a single laser pulse. An iris diaphragm (2) was used to isolate the central part of the beam, which had the least distortion of the wavefront. The laser pulse energy was regulated using a polarized attenuator (3) consisting of a half-wave plate and a Glan-Thomson prism. The energy control was performed using a calibrated germanium photo-diode (5), which registered the radiation reflected by a wedge-shaped glass plate (4). The sample (8) was mounted on a three-dimensional motorized stage (9) and was placed in the focus of lens (7) with a focal length of 200~mm. The $p$-polarized laser light was directed at the sample at an angle of $\sim 30^{\circ}$ to the normal of the target surface to avoid reflection back into laser. An optical microscopy scheme was assembled near the focusing lens, consisting of an Olympus PLN10$\times 0.25$ micro-lens (10) and a CCD camera (11), to select the exposure area on the sample surface. Magnification factor of the optical microscopy system equaled $M\approx 30$.

The spatial distribution of laser beam fluence on the sample surface has a Gaussian shape with the maximal fluence at the center of focal spot $F_0$ and the beam radius $r_0$ characterizing the beam size at 1/e$^2$ level. The parameters of $F_0$ and $r_0$ are determined according to the well-known method \cite{Liu1982} from the dependence of the square of ablation crater radius $r_a^2$ on the logarithm of the laser pulse energy $ln(E)$. According to this method, the sizes of a series of craters obtained at different laser pulse energies $E$ are measured in order to determine the Gaussian beam radius $r_0$. Since the oblique pulse incidence is implemented in the used scheme, each crater has an elliptical shape characterized by different sizes along the large $r_{0x}=85\un{\mu m}$ and small $r_{0x}=71\un{\mu m}$ axes. For the maximum pulse energy of $E = 1.3$~mJ applied in our experiments, the fluence in the beam center reaches $F_0=13.7\un{J/cm^2}$. Further energy increase results in the optical breakdown of the air near the sample surface.

\subsection{Methods of microstructural analysis}

Structural studies were carried out on a Tecnai Osiris transmission electron microscope (TEM) at an accelerating voltage of 200 kV. Thin foils (lamellae) cut perpendicular to the sample surface were prepared for TEM by ion thinning with a focused ion beam in the column of an FEI Scios scanning electron–ion microscope (Crystallography and Photonics Center for Collective Use). This method of preparing lamellae is described in detail in \cite {Montoya:2007}.

The lamellae were cut from the middle of separately located craters having the form of blurred and hardened local areas of the frozen melt formed after irradiation by a single laser pulse \cite{Khokhlov:2022}.
The protective coating material based on platinum was used to extract a thin lamella from a bulk sample. This is necessary to obtain an even transverse cut while maintaining thin subsurface layers directly adjacent to the surface. After deposition of the protective coating, a region up to $10\un{\mu m}$ wide along the working plane of the cut and up to  $7-8\un{\mu m}$ deep was cut out in the sample near the deposited platinum strip by an ion beam.
\begin{figure}[t]
\includegraphics[width=1.\columnwidth]{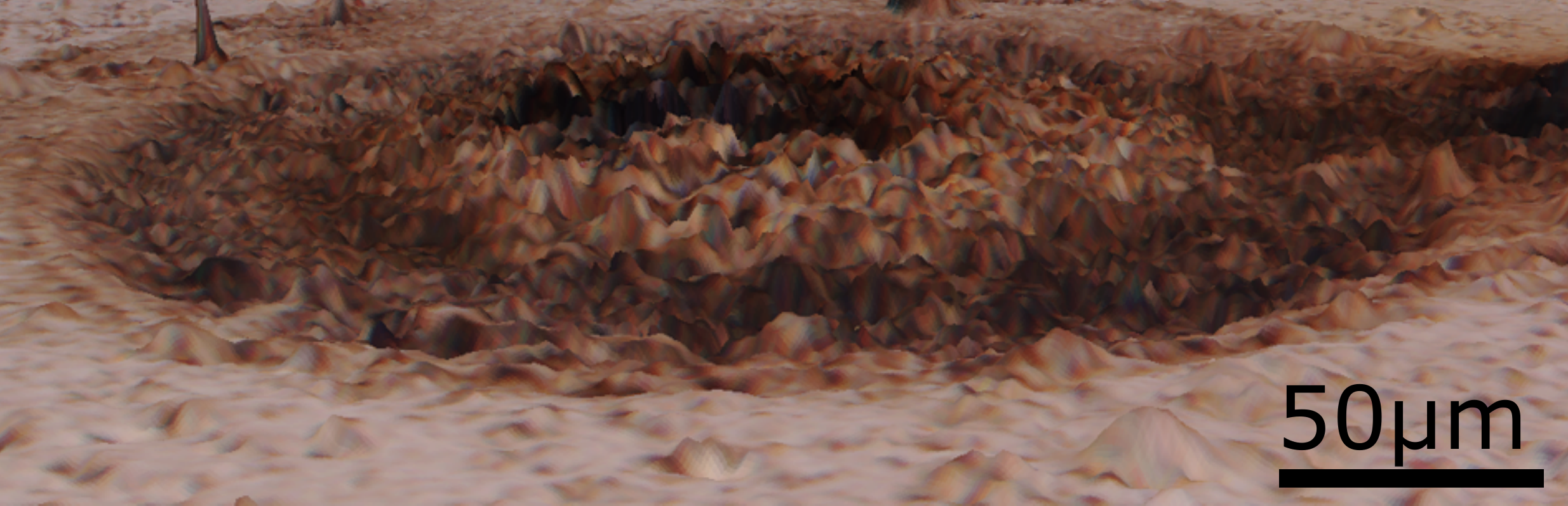}
\caption{\label{fig:02} General view of crater obtained by confocal scanning laser microscopy. The crater is formed on a surface of titanium sample irradiated by a powerful femtosecond laser pulse.}
\end{figure}
\begin{figure}[t]
\centering\includegraphics[width=1.\columnwidth]{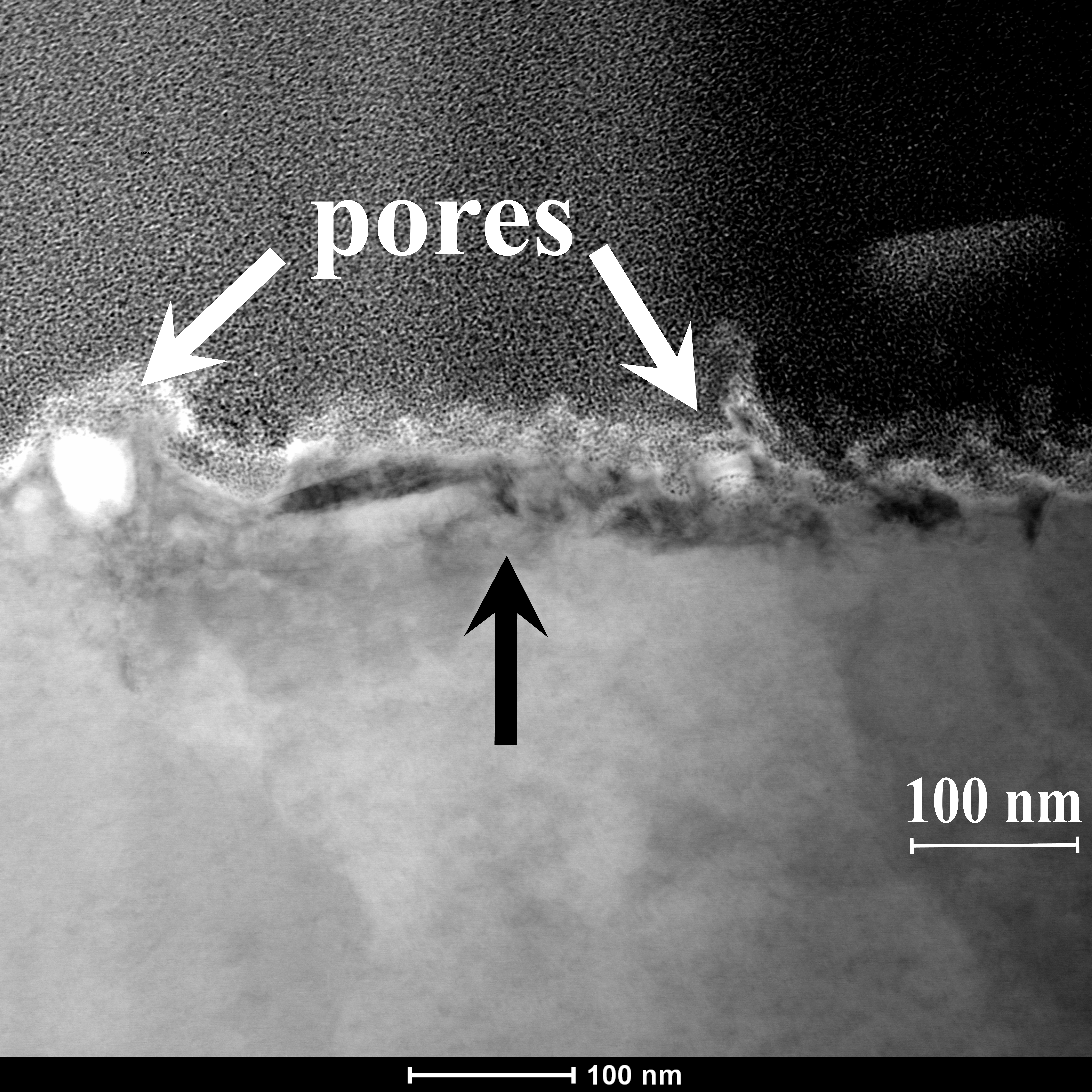}
\caption{\label{fig:pores} Transmission scanning electron microscopy of a lamella cut from the periphery of crater. The dark arrow indicates a modified surface layer about 100 nm thick with a chain of pores. }
\end{figure}

\begin{figure}[t]
\includegraphics[width=1.\columnwidth]{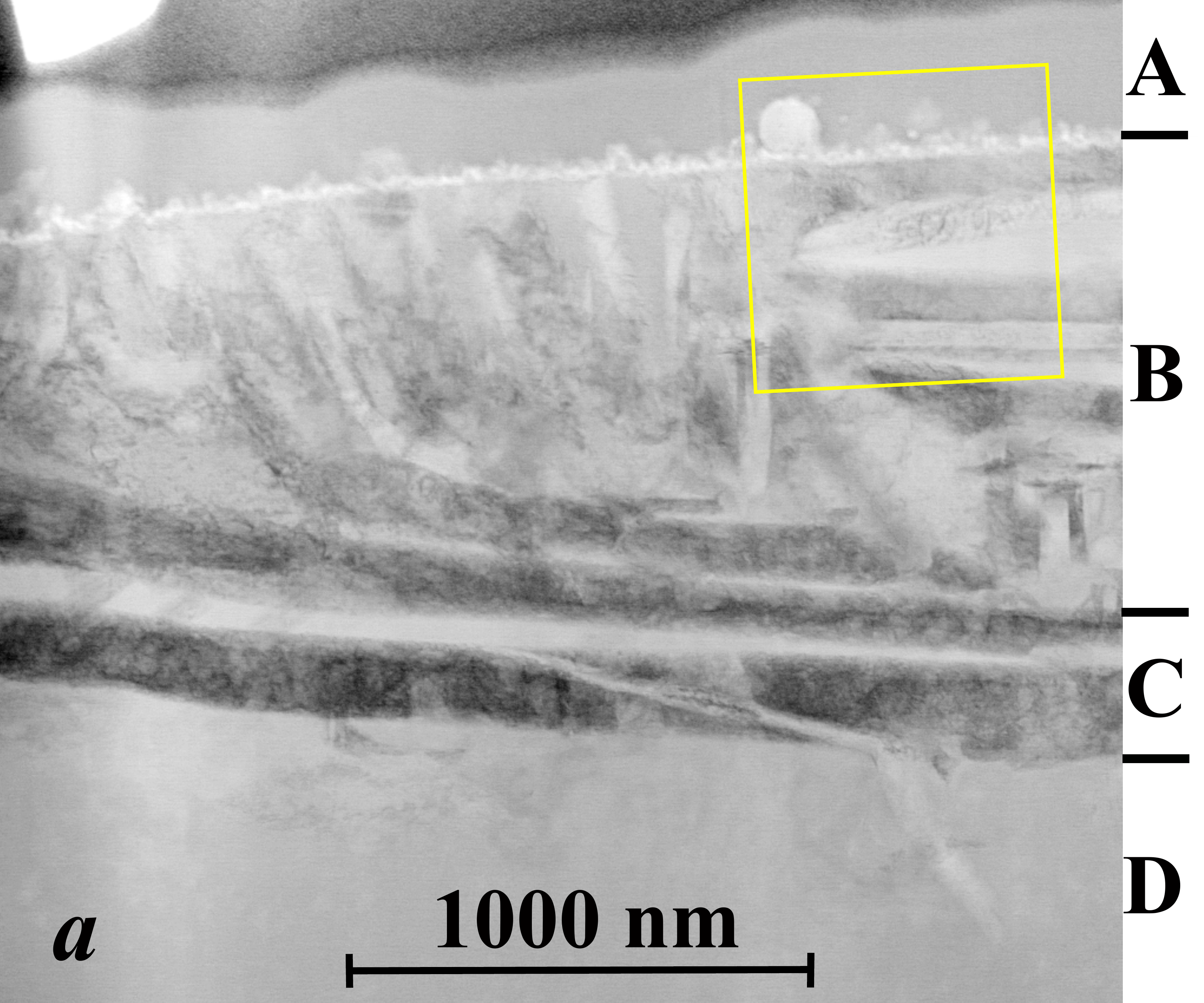}
\raggedright
\includegraphics[width=0.932\columnwidth]{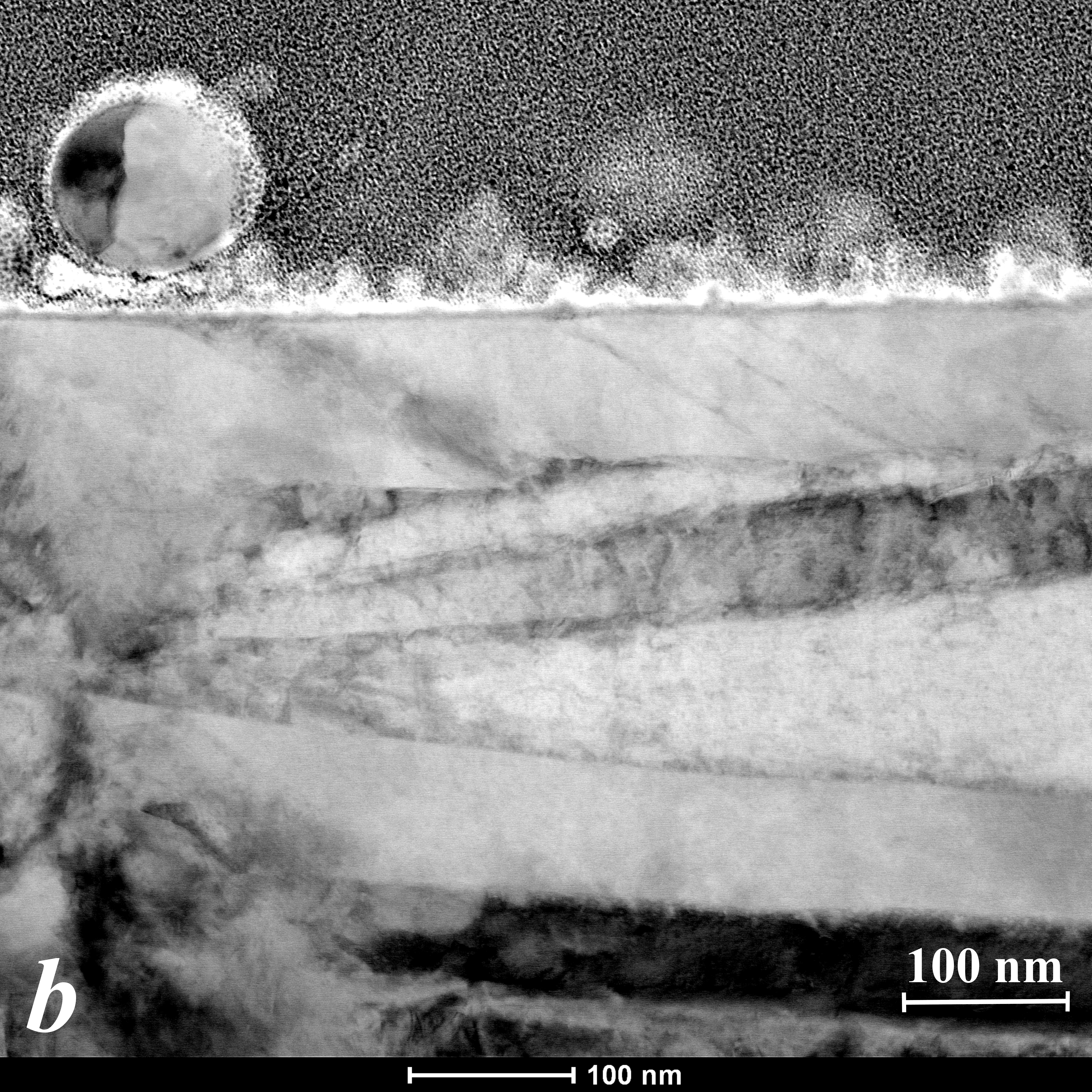}
\caption{\label{fig:plates} (a) Transmission scanning electron microscopy of a lamella with subsurface layers formed after laser pulse. The lamella was extracted from the central part of crater. The upper layer A is a platinum-based protective coating; the layer B corresponds to crystallization zone after SW melting of titanium; the layer C shows structures formed after plastic deformation and recrystallization induced by a weakening SW. Further attenuation of shock stress results in increase of dislocation density within the zone D, while the initial submicrocrystalline structure remains intact. The yellow square indicates area of enlarged image shown in (b), where the large frozen droplet of melt on the crater surface and the underlying plates in the layer B are clearly visible. }
\end{figure}

\section{Structure of near-surface titatnium layer after propagation of melting shock}
\label{sec:experimet}

A few lamellae were extracted from two regions of the craters obtained after laser irradiation. Some were taken from the periphery of craters nearby their rims, others were taken nearby the crater centers.  Figure ~\ref{fig:02} obtained with the confocal laser scanning microscopy shows the crater rim with oval edges formed after crystallization of molten surface.

With the help of transmission electron microscopy, one can see formation of porosity under the crater bottom in a near-surface layer of a lamella extracted nearby the crater rim, see Fig.~\ref{fig:pores}.
The pores are located at the depth of about 50-70 nm from the crater surface. The similar porous structure was formed during ultrafast cooling of a foam-like molten surface layer after irradiation of aluminum by a femtosecond laser pulse as reported earlier \cite{Ashitkov:2012:JETPLett}.

All following electron microscope images were obtained from lamellae extracted nearby the crater centers.
For the certain inclination angles of a examined thin lamella in the goniometer microscope those images show a modified subsurface layer approximately $1\un{\mu m}$ thick with a laminated structure, which is distinctly different from the original recrystallized state as seen in Fig.~\ref{fig:plates}(a). Frozen droplets of melt are also visible on the surface nearby the crater center. Those droplets are formed after formation of crater via ablation of heated surface material. According to elemental analysis, such globular particles (frozen droplets) are enriched with oxygen, as is the surface of sample itself. The enlarged part of this image nearby the biggest frozen droplet on the crater bottom is also presented in Fig.~\ref{fig:plates}(b).

The modified layer has a stratified grain-subgrain nanocrystalline structure with the plate widths from $\sim 5\un{nm}$ to $\sim 200\un{nm}$, and lengths from $\sim 100\un{nm}$ to several micrometers. It can be seen in the microstructure of the layer (B) shown in Fig.~\ref{fig:plates}(a,b) that the nano-sized plates are oriented in different directions in reference to the surface in the presented cross section. Some are perpendicular as it was pointed out in \cite{Khokhlov:2022}, where the subsurface layer melted by a strong SW was identified by the presence of such plates, and it was determined that the crystal lattices of adjacent plates were disoriented at angles of less than 10 degrees. Some others are oriented at the angle of 45 degrees or parallel to the crater surface. At the same time, the volume of material in the discussed layer (B) is divided into the individual plates including the plates parallel to each other. These plates are separated by high-angle grain boundaries.
In addition to the considered microstructure elements, the plate-shaped microstructure is also observed in the subsurface layer C, including those with a transverse size of several nanometers. The mechanisms of structure modification in shocked titanium in layers B and C are analyzed in the following sections \ref{sec:MDshock} and \ref{sec:MDtrans}.

\begin{figure*}
\centering
\includegraphics[width=0.75\textwidth]{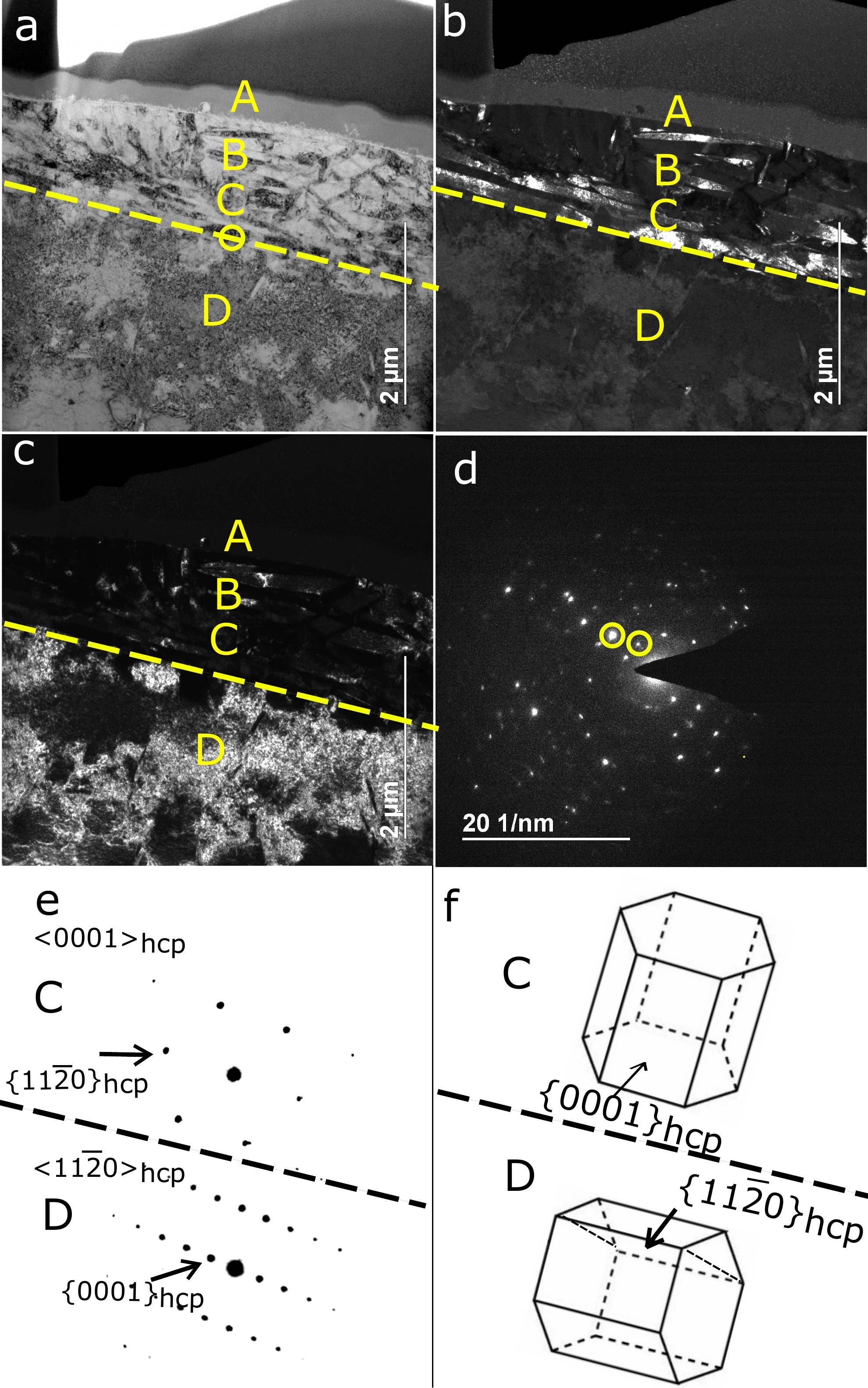}
\caption {Microstructure of SW-modified subsurface layers of titanium (B,C), as well as the deeper area (D), covered with a surface protective layer of metal-organic compound based on platinum (A) seen in Fig.~\ref{fig:pores} and \ref{fig:plates}: (a) -- light-field TEM image, where the yellow dotted line indicates the high-angle disorientation boundary between layers (C) and (D); (b,c) are dark-field images obtained in the reflexes marked in the microdiffraction pattern (d) by yellow circles, which belong to the diffraction patterns from the upper and lower regions on different sides from the high-angle boundary separating them: the left one is in the reflex of the main volume of the material (region D) from the family of planes $\{11\overline{2}0\}$, the right circle is in the reflection of the lamellar grain of the subsurface layer (region C) from the family $\{0001\}$; (d) - microdiffraction from the area of $0.4\un{\mu m}$ diameter (the circle marks the area cut by the selector aperture of size $0.4\un{\mu m}$ and capturing simultaneously zones (C) and (D); (e) is a scheme of microdiffraction patterns for grains in the layer (C) and (D) if they were oriented strictly in the axes of the $\langle 0001\rangle$ and  $\langle 11\overline{2}0\rangle$ type zones, respectively, (f) is a scheme explaining the approximate mutual orientation of grains between layers (C) and (D).}
\label{fig:hcp}
\end{figure*}

\begin{figure*}
\centering\includegraphics[width=1.\textwidth]{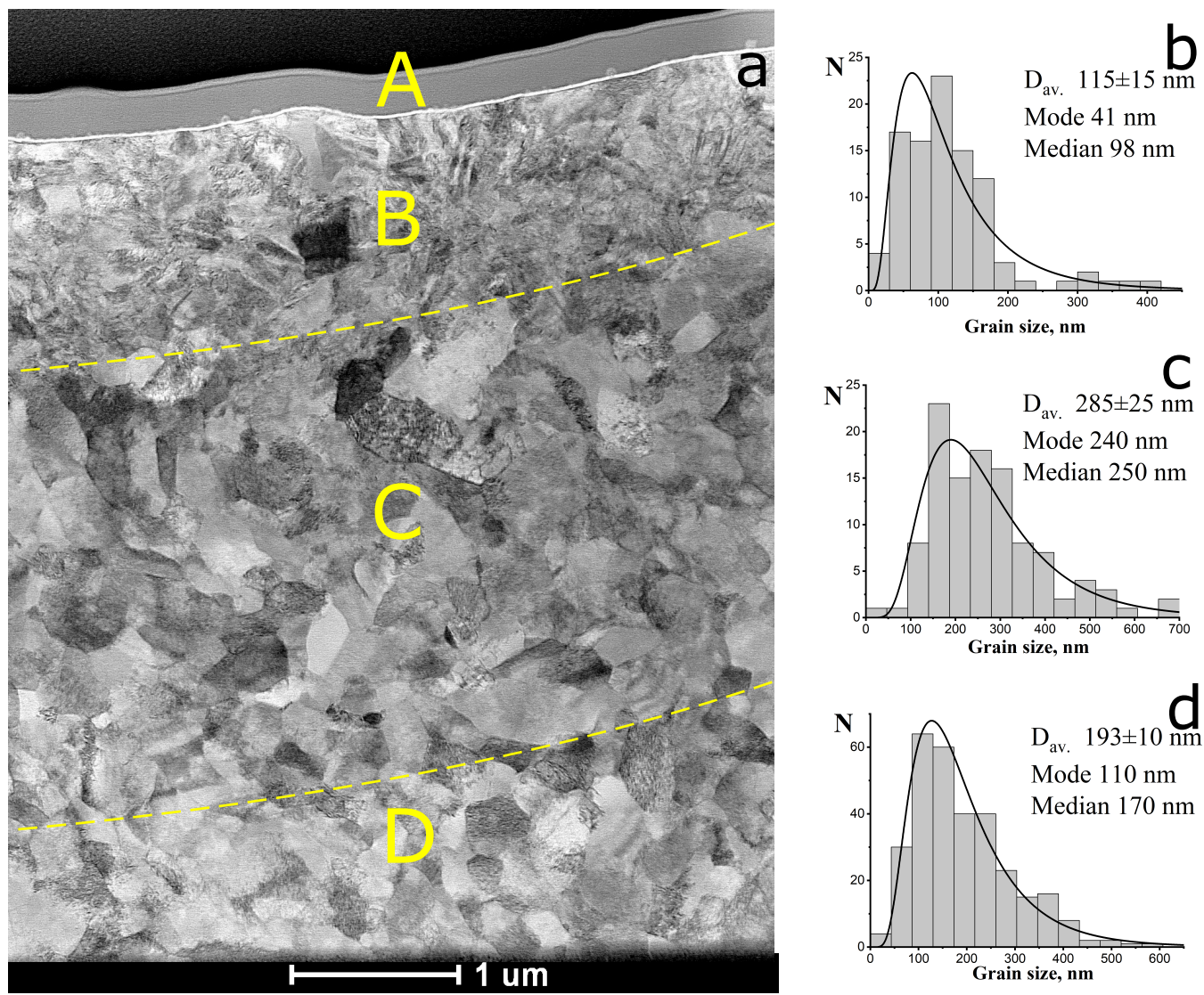}
\caption{Transmission scanning electron microscopy (a) of a lamella cut in the central part of crater on a sample initially consisting of small grains with the average size of 200 nm. A -- platinum-based protective coating,  B -- crystallization layer with nano-sized grains formed after SW melting and fast cooling, layer C corresponds to plasticity and recrystallization produced at lower shock pressure. D zone shows the initial submicrocrystalline structure remained intact after propagation of a sufficiently weakened SW. (b,c,d) -- histograms of grain size distributions for B,C,D layers, respectively.}
\label{fig:polycr}
\end{figure*}

 \begin{figure*}
\centering
\includegraphics[width=0.75\textwidth]{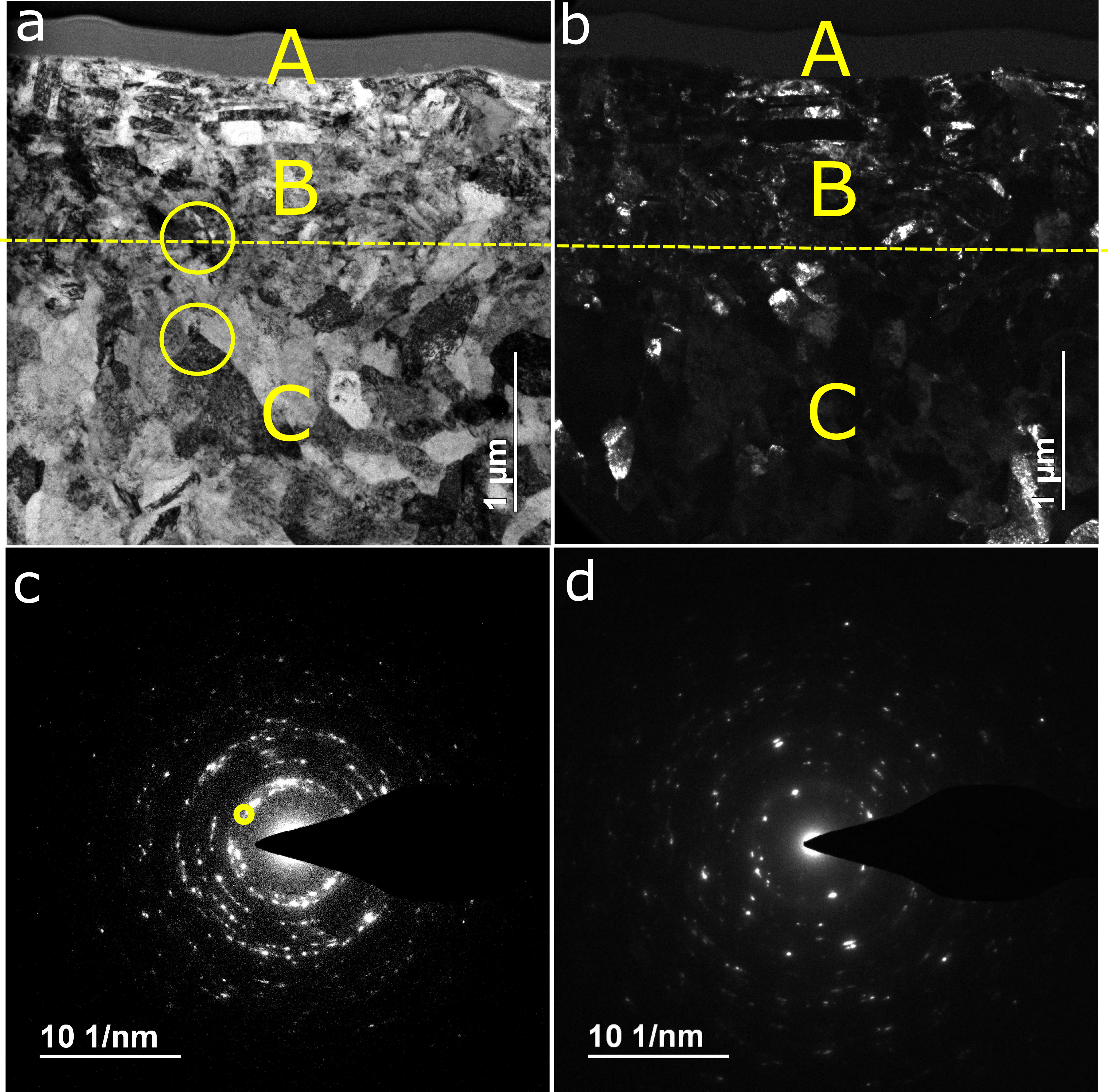}
\caption{Magnified TEM image of regions A, B and C presented in Fig.~\ref{fig:polycr}: (a) -- light-field image, where the yellow dotted line indicates the conventional boundary between regions B and C; (b) -- dark-field image obtained in the microdiffraction pattern marked reflection from the plane family $\{10\overline{1}0\}$ (yellow circle); (c) -- microdiffraction from a spot with the diameter of $0.4\un{\mu m}$ obtained from regions B and C, (d) -- from circles in region C shown in (a). }
\label{fig:07}
\end{figure*}

Other characteristic microstructure elements located below the considered layer (B) are the individual grains with lamellar shape observed in the light-field images. They are placed in the layer (C) shown in Fig.~\ref{fig:hcp}(b). The plates are oriented with their long sides predominantly parallel to the sample surface and are contiguous with the matrix (D), which is an initial material that has not undergone structure modification but has an increased density of dislocations. Microdiffraction analysis (Fig.~\ref{fig:hcp}(d)) and the dark-field images show that the misorientations between the crystal lattices of the parallel adjacent plates in the region (C) are high-angle (more than 15 degrees), see Fig.~\ref{fig:hcp}(c). The prismatic plane of family $\{11\overline{2}0\}$ in the crystal lattice within the region (D) is practically parallel to the basis plane of $\{0001\}$ type in the plate within the region (C) (the azimuthal angle between the planes is 13.8 degrees), i.e. the hexagonal crystals are practically perpendicular to one another.

In the zone (D) below the subsurface layer (C) the microstructure is characterized by a high dislocation density with distinctly pronounced formation of a cellular dislocation structure without visible misorientation between the adjacent cells, which indicates the absence of an excess of dislocations of the same sign in the cell walls. Such a cellular structure is observed over the entire area of the thin lamella under investigation, that is, at least up to the depth of $7-8\un{\mu m}$ where such layers with the thickness of a few microns are shown in ~\ref{fig:hcp}. The results of modeling the dislocation density dynamics in such layers after compression by a strong SW are given in Section \ref{sec:dislocation}.

It is known that the orientation of single-crystal sample along the propagation direction of an incident SW may change noticeably the response of material to uniaxial deformation including the cold melting as demonstrated in \cite{Budzevich:2012}. In this connection, it is of interest to investigate the subsurface melting in titanium samples of the same alloy as considered above, but in an initially submicrocrystalline state with the average grain-subgrain structure size of 200 nm at the same laser pulse parameters in order to compare with the results for coarse-grained titanium presented here in Figs.~\ref{fig:pores}--\ref{fig:hcp}. In such titanium with the grain size of about $30\un{\mu m}$ a whole lamellar area of up to $8\un{\mu m}$ wide observed in the transmission electron microscope must be a single crystal since the lamella size is smaller than a single grain.

Figure~\ref{fig:polycr} shows comparison of subsurface layers with submicrocrystalline structures formed in polycrystalline titanium after compression by a strong melting SW. In such a sample the numerous disoriented grain-subgrain elements are deformed by the SW front, and they should melt with the different rates. Thus, the melting front should be diffused. Indeed, the top subsurface layer B seen in Fig.~\ref{fig:polycr}(a,b) does not have a clear boundary with the layer C and comprises a nonuniform nanocristalline grain-subgrain structure of grains with the average size of about 100 nm, which are formed after the fast crystallization of melt. Below this layer at a depth of more than $1\un{\mu m}$ the second layer C (about $2\un{\mu m}$ thick) with the increased grain size up to 300 nm, apparently due to recrystallization of solid material,  is observed in Fig.~\ref{fig:polycr}(a,c). Beyond the depth of more than $3-4\un{\mu m}$ a region D with structure of intact submicrocrystalline titanium is observed, see Fig.~\ref{fig:polycr}(a,d). The microdiffraction picture obtained from a contact of regions B and C from a spot with diameter of $0.4\un{\mu m}$ (which captures the border areas of both regions) has a quasi-ring distribution of reflexes. This does not allow to clearly mark a boundary between the mentioned regions shown in Fig.~\ref{fig:07}(c). The microdiffraction image obtained in the region C from a spot with the diameter of $0.4\un{\mu m}$ has a noticeably smaller number of reflexes. This is due to a larger grain size in this region.

\section{Simulation of modifications in irradiated titanium}

\subsection{Shock-induced phase transitions in titanium}
\label{sec:MDshock}

\begin{figure}
\centering\includegraphics[width=1\columnwidth]{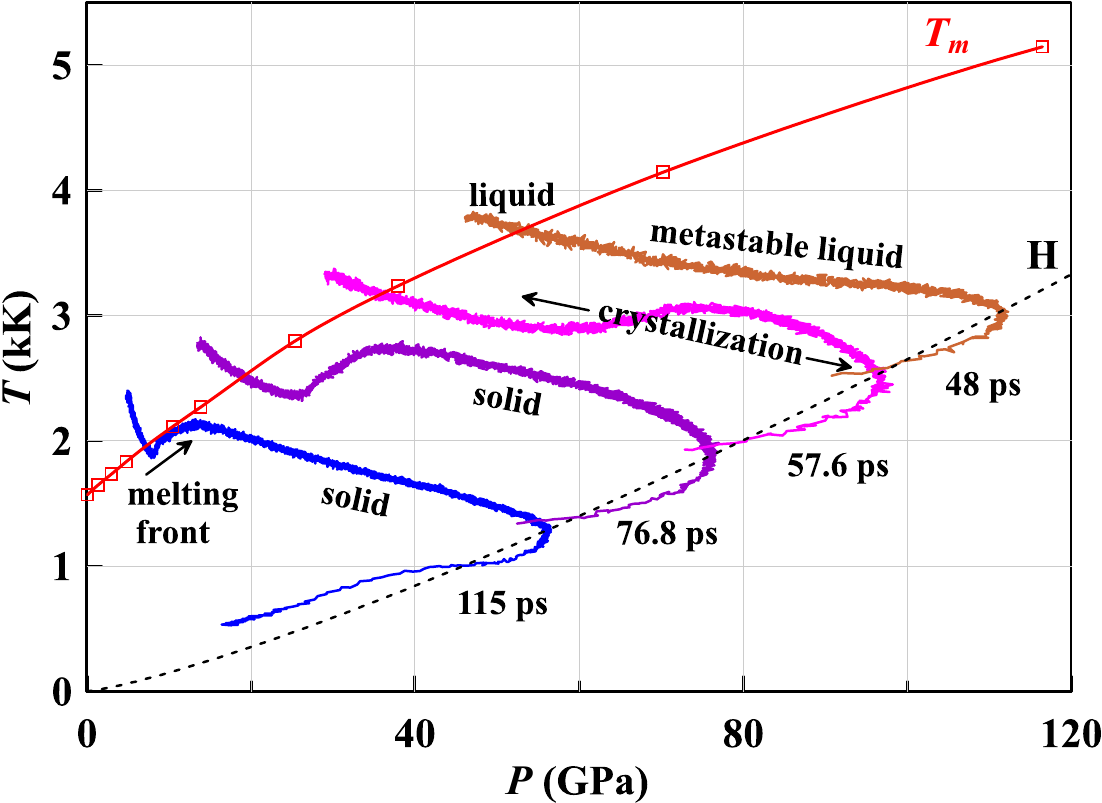}
\caption{\label{fig:T-P}
Distributions $T(P)$ from the spatial SW profiles $T(x)$ and $P(x)$ at different moments of time on the phase plane $T-P$. The melting curve $T_m(P)$ is derived from the results of MD modeling of equilibrium coexistence of liquid and solid titanium phases. The shock Hugoniot (H) connecting the pressure maxima in the SW front is shown by the dashed line. At early times $t<26\un{ps}$ the SW front creates enough compression to heat titanium above $T_m(P)$. The equilibrium melt states are labeled ``liquid''.
}
\end{figure}

\begin{figure}[t]
\centering\includegraphics[width=1\columnwidth]{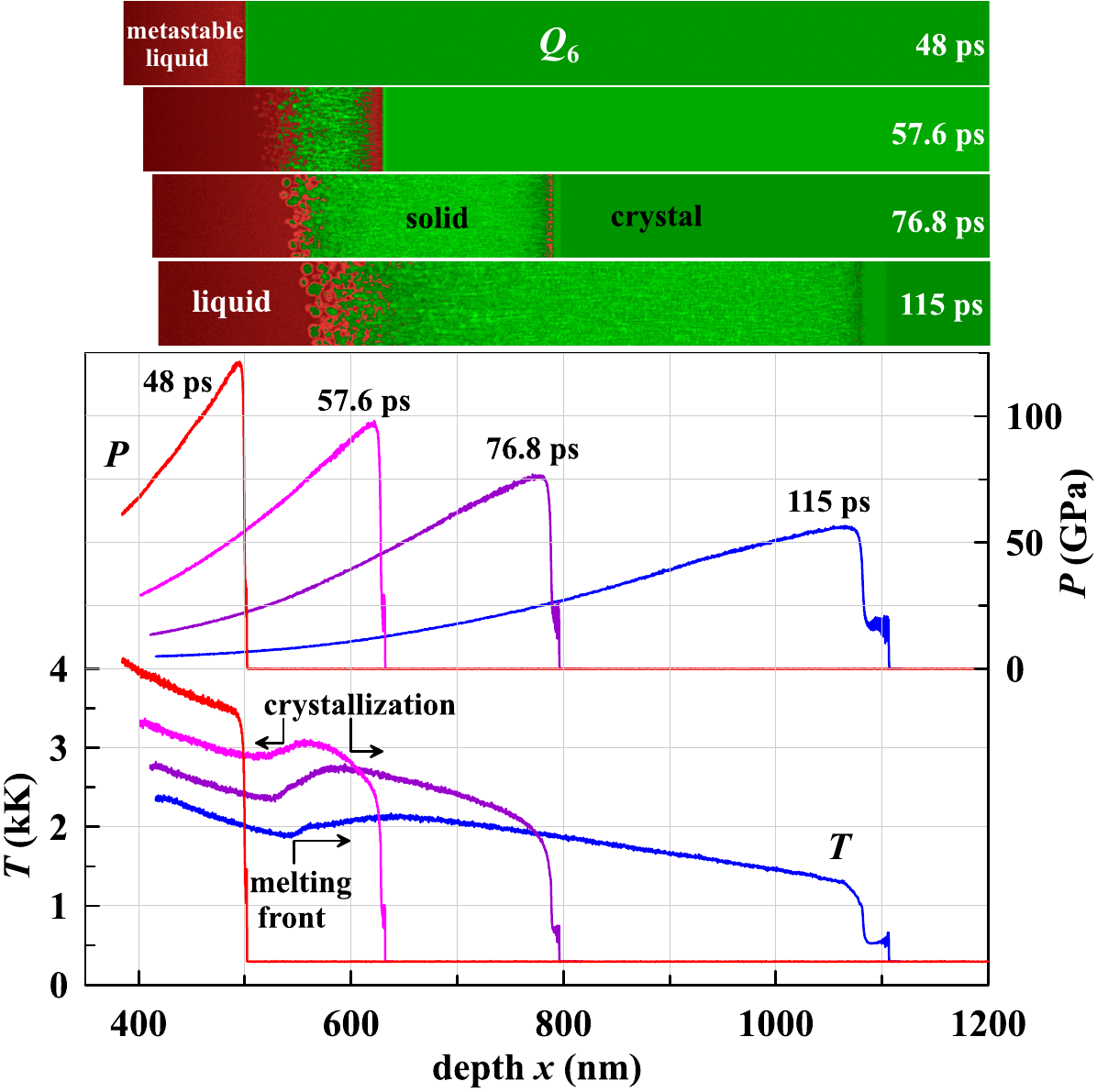}
\caption{\label{fig:PTQ_long}
Maps of the order parameter $Q_6(x,y)$, as well as the corresponding pressure $P(x)$ and temperature $T(x)$  profiles of SW propagating in titanium. After $30\un{ps}$ the SW weakens to $T_H < T_m(P)$ and the ``cold'' mechanical melting in the SW front begins producing a metastable supercooled melt.
Degree of supercooling in melt increases as the SW weakens, which leads to rapid crystallization just after the SW front, see $Q_6$ maps at $t=[57.6 - 76.8]\un{ps}$. After about $72\un{ps}$, the cold melting stops due to the weakening of SW below $P\sim 80\un{GPa}$. After $110\un{ps}$ the metastable melt solidifies entirely, but the equilibrium melting begins as a solid phase acquires temperature above $T_m(P)$ during unloading. The profiles shown are also represented in the $T-P$ plane in Fig.~\ref{fig:T-P}.
}
\end{figure}

We applied a widely used hydrodynamic method to simulate a nonequilibrium two-temperature (2T) electron-ion system \cite{Anisimov:1974} in order to determine an amplitude of a shock wave generated within a surface layer of titanium during near isochoric heating of a lattice by the hot electrons heated by a femtosecond laser pulse. The used system of Lagrangian 2T hydrodynamics (2T-HD) equations was given in \cite{Anisimov:2019:JETP}. To model titanium the wide-range equation of state \cite{Povarnitsyn:2008,Khishchenko:2016} is employed.

Absorption of a powerful 110 fs laser pulse taken for our simulation results in energy deposition of $5.5\un{J/cm^2}$, which is by two orders of magnitude greater than the ablation threshold of titanium \cite{Ashitkov_Titanium_ablation-1,Ashitkov_Titan_spallation-5}. The electrons in an absorbing skin layer of $\delta\approx 12\un{nm}$ acquire the temperature of $T_e\sim 10\un{eV}$ while the lattice (ion) temperature $T_i$ remains relatively cold during the pulse. 2T stage of energy transfer from hot electrons to ions lasts about 3 picoseconds during which a heated layer goes deep into a sample to a depth of $d_T\approx 30\un{nm}$ by electron heat conduction. As a result this heated layer melts and ablates. After the 2T stage the thickness of heated layer increases with a damped speed which is slower than the sound velocity.
Immediately after electron heating, the electron pressure reaches $P_e\approx 1000 \un{GPa}$, and just after the 2T stage the total (electron plus ion) pressure drops to $P=P_e+P_i\approx 450 \un{GPa}$. A shock wave of such high amplitude propagates deep into cold titanium with speed of $\sim 15 \un{km/s}$, and in doing so heats and melts metal in the SW front until the pressure in the front is sufficiently relaxed.

From the solution of the 2T-HD equations, the velocity of a Lagrangian particle (LP) with a position deeper than an increasing thickness of heated layer at a time of SW arrival is determined. For this purpose, a Lagrangian particle at the initial depth of 300 nm is chosen, since it does not have time to warm up appreciably due to thermal conductivity before the arrival of SW. The recorded LP velocity was then used to set the piston velocity in molecular dynamics (MD) simulation of SW propagation in titanium beyond the depth of 300 nm. The LP velocity reaches $4\un{km/s}$ abruptly at 23 ps,  then falls smoothly to $2\un{km/s}$ at 36 ps, and decays to 0 during next 100 ps.
The details of 2T-HD simulation of titanium, as well as the transfer of information to the MD method in order to continue smoothly the MD simulation of melting and crystallization processes at the atomistic level in an unloading tail behind the SW front during its attenuation, are described in detail in \cite{Khokhlov:2022}.

For the MD simulation of phase transformations in titanium under compression by such a powerful SW, we applied the EAM potential \cite{Zhou:2004} satisfactorily describing the shock Hugoniot and melting of titanium. The melting curve $T_m(P)$ shown in Fig.~\ref{fig:T-P} is obtained from MD simulations of the equilibrium phase coexistence of liquid and solid phases at different pressures using a small simualtion daomain. The intersection of the shock Hugoniot and melting curves occurs at $P\approx 170\un{GPa}$ and $T\approx 5.84\un{kK}$. As we will show below, a weaker SW can also melt the material via its transformation to a metastable liquid state.

To simulate SW propagation the elongated titanium samples of $1000\times 20 \times 20\un{nm^3}$ and $1000\times 80 \times 20\un{nm^3}$ were created and thermalized at $293\un{K}$ and zero pressure. Those samples with different orientations $[0001]$, $[01\overline{1}0]$, and $[2\overline{11}0]$ are studied, but only the latter one is discussed in detail here since a shocked crystalline grain had fortunately the same orientation in our experiment as seen in Fig.~\ref{fig:hcp}(f).
It is also worth noting that the pattern of SW melting and crystallization of titanium is almost independent on orientation and cross-sectional area of the samples considered.

The initial density of titanium is $4.393\un{g/cm^3}$ and the total number of atoms in the simulations was approximately $89\times 10^6$.
Periodic boundary conditions are applied only in the transverse directions on the $YZ$ plane. A stiff repulsive potential is used as a piston at the left boundary of the sample at 300 nm depth on the $X$ axis It corresponds to the initial position of the moving LF. The right boundary at 1300 nm depth is completely free (no evaporation of atoms into vacuum at the room temperature). Because of the shock compression and high temperature in the shock front, the integration step is chosen small enough to be $0.5\un{fs}$. The MD simulation of SW propagation is performed with our in-house parallel code MD-$\mathrm{VD^3}$ utilizing the Voronoi dynamic domain decomposition with highly efficient computational load balancing algorithm \cite{Zhakhovsky:2005,Egorova:2019}.

\begin{figure}[t]
\centering\includegraphics[width=1\columnwidth]{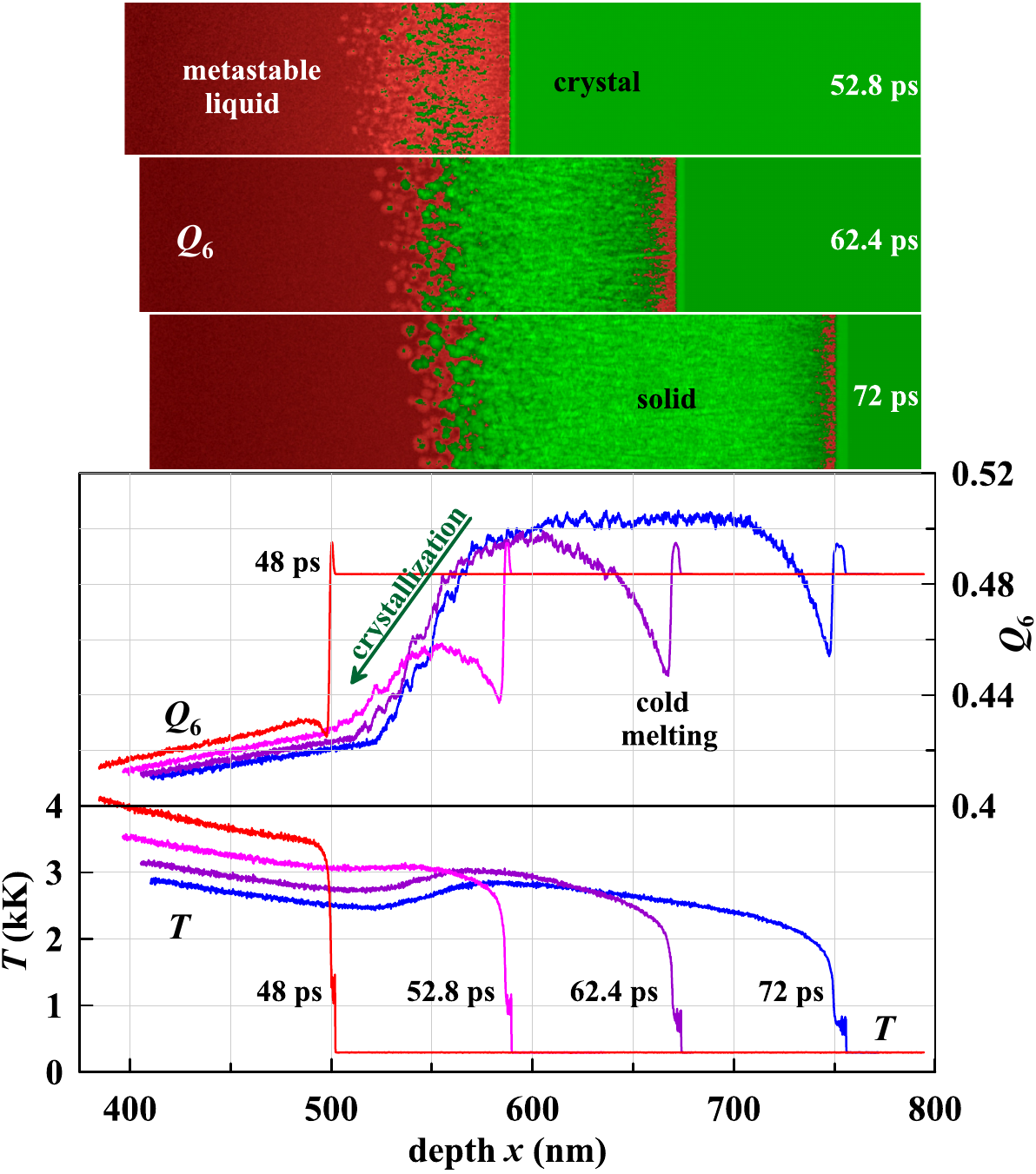}
\caption{\label{fig:TQ_short}
Maps of the order parameter $Q_6(x,y)$, as well as the corresponding profiles of $Q_6(x)$ and temperature $T(x)$, averaged over the $YZ$ cross section, obtained from MD simulation of SW propagation in titanium. Crystallization of the metastable melt begins with nucleation of the solid phase at 52.8 ps.  Two crystallization fronts are formed -- the right front quickly catches up with the SW front, and the left front moves  slowly in the opposite direction to the SW front in the hotter melt, where solid grain growth can be seen. See Supplemental Movie at [URL will be inserted by publisher] for for more details.}
\end{figure}

The shock wave is transmitted from the 2T-HD to the MD simulation using the LP velocity recorded in 2T-HD modeling, which is used to set the velocity of a piston simulated with a hard repulsive potential. By 23 ps, the SW has weakened markedly and enters the MD sample at speed of $10.2\un{km/s}$ -- with the maximum velocity of the piston, and therefore the material, reaching $4\un{km/s}$, and pressure in the SW front jumping $185\un{GPa}$. However, this shock compression is still sufficient to heat titanium above the melting curve $T_m(P)$. For the first time, the amplitude of the SW front drops below $T_m(P)$ at $t=30\un{ps}$, but the melting does not stop, but instead enters a regime of ''cold'' mechanical melting in the SW front where formation of metastable supercooled liquid \cite{Budzevich:2012} happens as shown in Figs.~\ref{fig:PTQ_long} and \ref{fig:TQ_short}.

In addition to the temperature profiles, there are the 1D and 2D distributions of the local atomic order parameter $Q_6(x,y,t)$ \cite{Steinhardt:1983,Zhakhovsky:2017} presented to illustrate the solid-liquid phase transition. Also, the spatial pressure and temperature distributions shown in Fig.~\ref{fig:PTQ_long} are presented in the $T-P$ plane in Fig.~\ref{fig:T-P} to determine the phase state of shock-compressed titanium with respect to the melting curve $T_m(P)$ and the shock Hugoniot $T_H(P)$. In particular, the spatial profiles of $T(x)$ and $P(x)$ at $t=48\un{ps}$ are plotted as one $T(P)$ curve in Fig.~\ref{fig:T-P}. The distribution $Q_6(x,y)$ at $t=48\un{ps}$ in Fig.~\ref{fig:PTQ_long} indicates that titanium behind the SW front is entirely in a molten state. At the same time, the phase diagram in the $T-P$ plane shows that almost all of the melt lies in the area of equilibrium solid states below the melting curve, except for titanium near the piston which was melted at an early times $t<30\un{ps}$, when its temperature was $T_H > T_m(P)$ since the SW had not yet weakened enough. Thus, almost the entire melt at $t=48\un{ps}$ is in a metastable supercooled state.

Cold melting completely stops after the weakening of the SW below the threshold pressure, at which the maximum shear stress in the front falls below $\tau \sim 12 \un{GPa}$, which is close to the threshold of free slipping of atomic planes at the critical Frenkel stress $\tau^* \approx G/(2\pi)$ \cite{Hull:2011}, which for uncompressed titanium with the shear modulus of $G=44\un{GPa}$ is $\tau^* \approx 7\un{GPa}$. It should be noted that the cold melting during shock compression along some axis of a single crystal may not develop if the shear stress achieved in the SW front is less than the critical stress. For example, such a situation can occur in FCC lattices during compression exclusively along the $\langle 100\rangle$ family of axes \cite{Budzevich:2012}. In polycrystalline materials, the rare grains may be found in such orientations with respect to the compression axis. In our simulations of single-crystal samples having $[0001]$, $[01\overline{1}0]$ and $[2\overline{11}0]$ orientations, the maximal shear stresses within the SW fronts are slightly different while the maximal longitudinal stresses are identical. Therefore the cold melting stops when the SWs propagating in those directions weaken to the different pressures of $P\sim 71,\; 81,\; 80 \un{GPa}$, respectively. The resulting melting depths are 890, 720, and 760 nm for corresponding crystal orientations at the same absorbed laser pulse energy.

As the SW weakens, the temperature at the SW front drops and the degree of supercooling increases causing the produced metastable melt to crystallize in the unloading tail behind the SW front, see maps $Q_6(x,y)$ for times $t\in [52-77]\un{ps}$ in Fig.~\ref{fig:PTQ_long} and \ref{fig:TQ_short}.
Crystallization of the melt begins with formation and growth of solid phase nuclei at 52 ps, shown in green coloring on map $Q_6(x,y)$ in Fig.~\ref{fig:TQ_short}. The emergence of those small nuclei causes a local increase in $Q_6(x)$ and temperature $T(x)$, as seen in their spatial profiles. After formation of a zone with solid phase behind the SW front, one can assert that two crystallization fronts are formed and they propagate with notably different velocities in opposite directions in the metastable melt.
The fastest crystallization occurs in the right front catching up with the SW front, while a slower one develops in the hotter unloading tail of SW, leading to a gradual growth of solid grains in the left crystallization front, as seen in Fig.~\ref{fig:PTQ_long} and \ref{fig:TQ_short}, where the directions of motion (in the local coordinate system of moving material) of two corresponding crystallization fronts are indicated -- also see Supplemental Movie at [URL will be inserted by publisher].

Propagation of those crystallization fronts in the metastable melts with very different supercooling degrees results in different crystalline structures formed behind the fronts. Mechanical melting in the SW front leaves many tiny nuclei of a crystal phase in the formed melt. Just after the SW front such nuclei begin to grow predominantly in direction of lower temperature (towards the SW front), and soon these nanosized crystallites coalesce into the fastest crystallization front. As a result, the solid after this front consists of crystallites elongated in parallel to direction of shock propagation. By contrast, the left crystallization front propagates in a hotter melt formed at higher shear stress and temperature, where number of remaining solid nuclei is lower and their grow rate is much smaller. It results in direction-independent growth of crystalline grains.
Such patterns are visible clearly after 60 ps in Fig.~\ref{fig:TQ_short} and Supplemental Movie at [URL will be inserted by publisher].

The SW accelerates material to the right, and the slowest left crystallization front stands almost at rest in the computational domain, as seen in the maps of atomic order parameter in Fig.~\ref{fig:TQ_short}. Therefore, this front moves leftward through the material at a local speed of $200\un{m/s}$ between 62.4 and 72 ps.
Crystallization in this front develops up to about 110 ps, when the local temperature goes above the melting curve $T_m(P)$ in response to the decrease of pressure, which puts the metastable melt into a thermodynamically equilibrium liquid state, as shown in the $T(P)$ profile at $t=115 \un{ps}$ presented in Fig.~\ref{fig:T-P}. It can also be seen from this profile that further unloading in the SW tail transfers the already solidified titanium above the $T_m(P)$, causing it to melt and leading to formation of a new melting front moving to the right, as indicated in Fig.~\ref{fig:T-P} and \ref{fig:PTQ_long}. Shortly after this time, the simulated SW reaches the right edge of sample and is reflected as an unloading wave with a large tensile stress amplitude. Propagation of such unloading wave in the left part of sample results in an additional stretching which changes greatly the long-term evolution of material, in particular freezing and recrystallization of titanium, and also causes spallation at the right end of the sample. All those phenomena are not considered here since they are not happened in the experimental conditions.

\begin{figure}[t]
\centering\includegraphics[width=1.0\columnwidth]{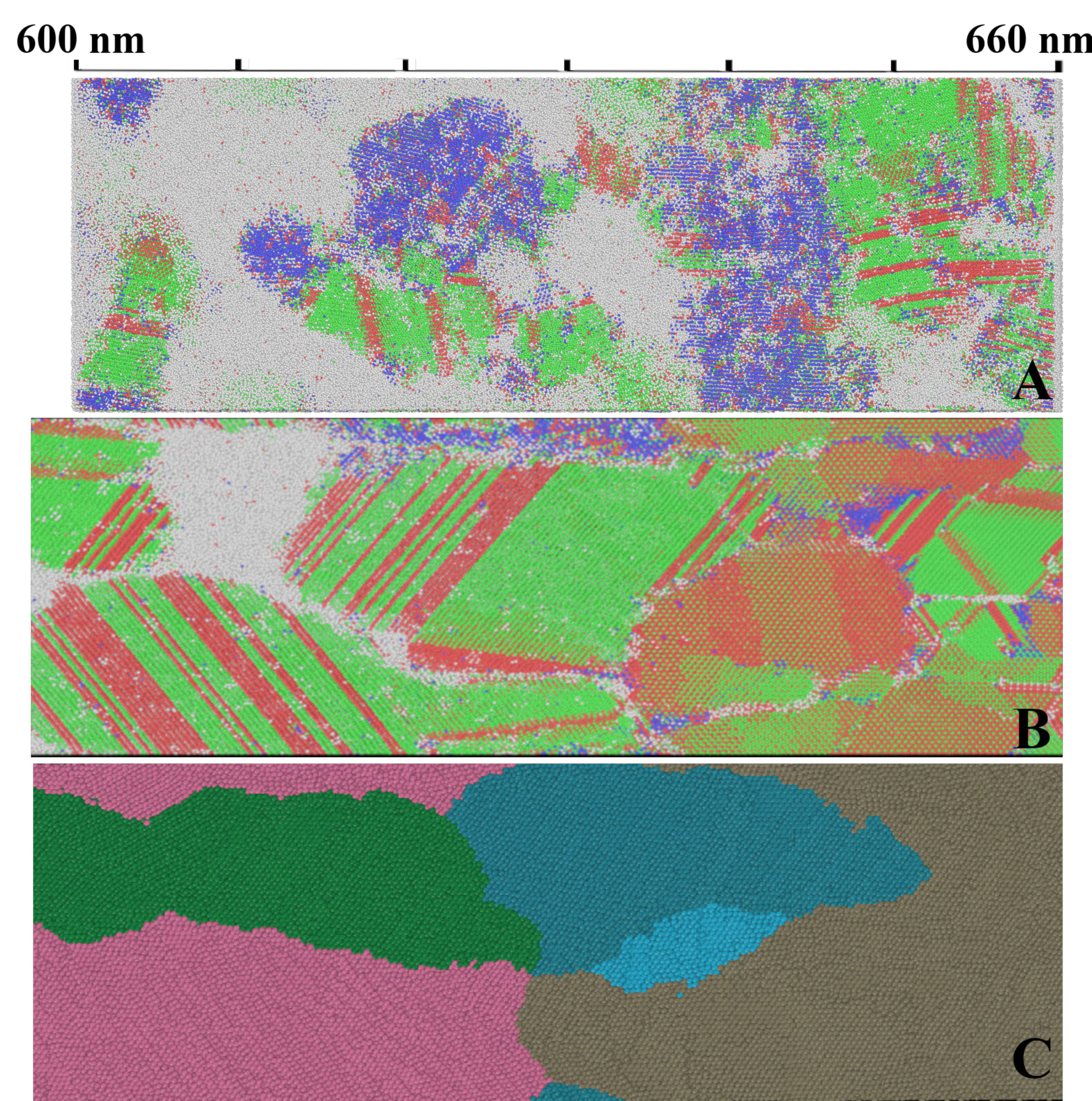}
\caption{\label{fig:mel} Atomic structure of the melt-solid interface in a small sample: (A) -- extracted just after SW passage at $t=0.131\un{ns}$, (B) -- the same sample at $t=200\un{ns}$ after long-term cooling and pressure relaxation. Atoms with disordered neighbourhood like in  melt are shown in white, while atoms with surrounding atom positions typical for FCC, BCC, and HCP structures are shown in red, green, and blue, respectively. (C) the same sample after completed crystallization at $t=250\un{ns}$. As a result, a polycrystalline structure with grains oriented in parallel with the direction of maximal heat flux is formed. Segmentation into grains (marked with different colors) is determined in the Ovito program. Top ruler indicates the sample position below the heated surface.}
\end{figure}

\subsection{Long-term evolution of atomic structures after shock}
\label{sec:MDtrans}

To study the long-term evolution of the atomic structures obtained after large-scale MD simulation of the SW propagation and shock-induced transformation within the SW front and in its unloading tail, it is necessary to perform pressure relaxation and temperature reduction with heat dissipation directed deep into the material. For small-scale MD simulation of such slow directional crystallization process, the small part containing the solid/melt boundary was extracted from the large sample (with a smaller cross section). Extracting the small sample with dimensions $65\times 20\times 20\un{nm}$  also avoids the impact on the phase state of material by the tensile wave reflected from the rear-side boundary in the large sample. A view of the small sample before crystallization is shown in Fig.~\ref{fig:mel}(A).

The LAMMPS software package \cite{LAMMPS} and the same titanium potential applied for the large sample are used to simulate material transformation at the melt/solid zone in the extracted small sample. The evolution of atomic structure in this sample is simulated for 250 ns with an integration step of $2\un{fs}$.
Periodic boundary conditions are imposed only in the YZ section, while the potential walls are applied at the sample boundaries on the longest X-axis. Relaxation of the system to zero pressure is performed independently for each axis.

The pressure relaxation rate $50\un{MPa/ps}$ is taken from the SW unloading tail at the end of large sample simulation discussed in Section \ref{sec:MDshock}, and this relaxation is applied until $P=0.$
The heat removal from a zone of 5 nm thick at the farthermost edge of sample on the longest axis, provides an initial temperature decrease rate in this zone of $0.47\un{K/ps}$, which decreases as the sample is cooled to the target temperature of 1100 K.
According to \cite{JIANG2021138187} this rather slow cooling process makes it possible to trace evolution of phase composition within 250 ns, which ends with formation of the crystal structures. By the end of MD simulation, the temperature of this sample settled at about 1100K.

To analyze the initial structure, the methodology proposed in \cite{Larsen:2016robust} is used with the established standard deviation from the ideal structure template RMSD=0.1. Figure~\ref{fig:mel}(a) is plotted by the Ovito software \cite{Stukowski:2009}, where atoms with the local environment corresponding to the melt are marked in white, and atoms with the HCP, FCC and BCC lattice environments are marked in red, green and blue, respectively.
Our MD simulation reveals the following characteristic features in formation of atomic structures in the region containing the melt/solid body boundary. If the heat sink is directed than the columnar structures are formed with their elements elongated along the direction of the maximum heat transfer through the solid part of the sample.

There are competing processes are happened -- melting of the particles surrounded by the melt and grain growth starting from the melt/solid boundary. Disorientation of the forming grains is promoted by the presence of crystallization nuclei at the melt/solid boundary, which are disoriented titanium clusters -- the grain segmentation analysis \cite{bonald2018hierarchical}) performed in Ovito software is presented in Fig.~\ref{fig:mel}(b).
The overall pattern of the columnar structure correlates with that observed experimentally, see Fig.~\ref{fig:plates}.

Long-term relaxation of titanium subjected to plastic deformation by an attenuated shock wave was also performed in an additional MD simulation. A small sample of $20\times 20\times 20 \un{nm}$ was extracted from a region near the right border of the large small-section sample. The total simulation time of the relaxation process to zero pressure takes about 4.2 ns.

We found that two grains extending parallel to the sample surface are formed with a $\sim 59^\circ$ misorientation angle at their boundary. The volume of the formed grains having HCP crystal structure is saturated with dislocations and packing defects with a fraction belonging to these defects of atoms of the order of 20\%. The relaxation process is accompanied by a decrease in pressure to zero, and the fraction of packing defects is reduced to 13\% due to the displacement of layers of atoms along the defects on the grain boundaries.

\begin{figure}[t]
\centering\includegraphics[width=1.0\columnwidth]{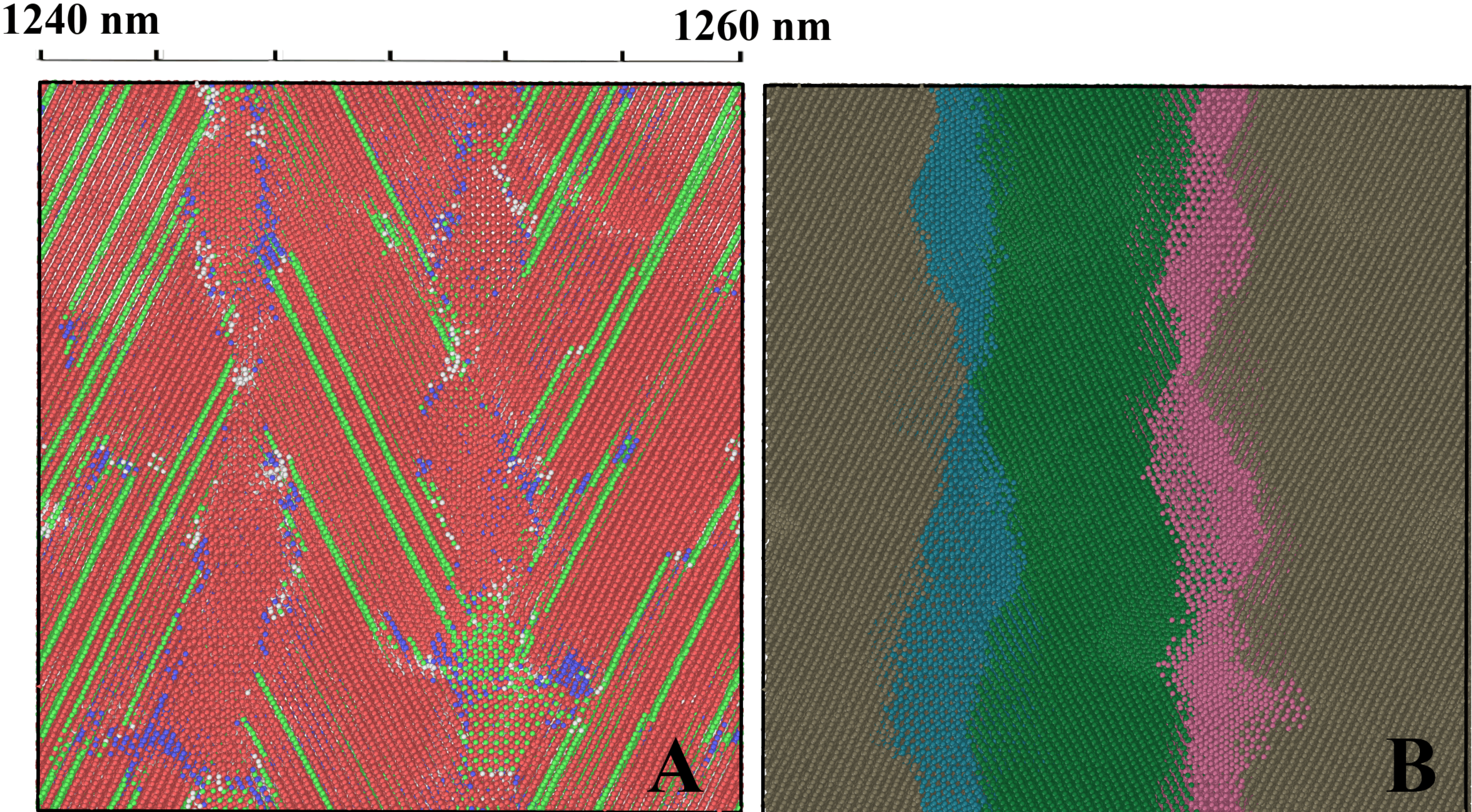}
\caption{\label{fig:me2} Structure of the plastic deformation region in a small sample after SW propagated from the left to right: (A) -- distribution of the local atomic order after long-duration relaxation: red, green, and blue are colored by atoms with local surroundings characteristic of HCP, FCC, and BCC structures, respectively; white indicates disordered atoms at grain boundaries, (B) -- segmentation into grains of different orientations marked with different colors determined in the Ovito. Top ruler indicates the sample position below the heated surface.}
\end{figure}

An illustration of the structure formation process is shown in Fig.~\ref{fig:me2}(a), where the surface of the original large sample is located far beyond from the left boundary. The rows of atoms with FCC structure (colored by green) surrounded by atoms with HCP structure (red color) represent the packing defects. Figure~\ref{fig:me2}(b) obtained by the grain segmentation analysis highlights the grains with thickness of the order of 10 nm formed in the deformation zone C of titanium after the shock-wave compression. The thickness of these grains is expected to increase as the width of the extracted small samples increases in the direction towards the original surface.

On the basis of the data obtained in the present work and our previous results \cite{Nelasov_2021} we can conclude about three possible mechanisms of structure refinement, which are activated if the sample surface is exposed to a femtosecond laser pulse --- the directed crystallization, shock-induced deformation, and complex process including phase recrystallization through formation of high-pressure phase along $\alpha-\omega-\alpha$ path and through the high-temperature $\beta$-phase titanium.

Martensitic transformation of $\alpha$ to $\beta$ phase, according to previous studies of thermodynamic equilibrium conditions between solid phases \cite{PhysRevB.78.054121,KARTAMYSHEV201930}, must happen in solid at a high temperature, that is nearby a solid-liquid interface, after decrease of pressure in the unloading tail of SW. Thus, formation of $\beta$ phase occurs predominately in parallel to this interface and the target surface, which results in elongation of grains. Such elongated grains are experimentally observed in area C as shown in Fig.~\ref{fig:plates} and ~\ref{fig:hcp}, that argues for the  $\alpha-\beta$ transition suggested above. A detailed investigation of the mentioned phase transformation processes requires a further MD simulation with an improved interatomic potential constructed within the framework of modified embedded atom method in order to reproduce more accurately the phase transitions in titanium \cite{PhysRevB.78.054121,KARTAMYSHEV201930,KIM20083481}.

\subsection{Formation of cellular dislocation structures}
\label{sec:dislocation}

When a high-pressure shock compression is applied to solid, the dislocations are produced within individual crystalline grains and density of dislocations increases there. If the dislocation density becomes high enough then the cellular dislocation structures are formed, as discussed in \cite{Estrin:1998}. To describe the formation of such structures, we performed a one-dimensional continual modeling of shocked titanium by taking into account an elastic-plastic model guided by kinetics of dislocations. The parameters of this model, which control the plastic response of the material, are determined by the dislocation density. The dislocation kinetics in the one-dimensional system of equations of continuum mechanics in the Lagrangian coordinates is written as in \cite{Krasnikov:2010}:
\begin{eqnarray}
\frac{dS_{zz}}{dt}&=&\frac{4}{3}G\frac{\partial u}{\partial z}+\frac{2}{\sqrt{6}} G V b n_d,\\
\frac{m_0}{(1-V^2/c^2_t)^{3/2}}\frac{dV}{dt}&=&-\frac{3b}{4} \left(  \sqrt{\frac{2}{3}}S_{zz}\pm \frac{2}{3}Y \right) - \mu(V,T)V \nonumber
\end{eqnarray}
Here the standard notations are used: $\rho$ -- density, $u$ -- mass velocity, $\sigma_{zz}$ -- stress tensor, $P$ -- pressure, $S_{zz} = \sigma_{zz} + P(\rho,E)$ -- stress tensor deviator, $Y$ -- yield strength, $b$ -- Buergers vector, $T$ -- temperature, $V$ -- dislocation velocity, $\mu$ -- dislocation inhibition coefficient, and $n_d$ -- dislocation density.
The following equation \cite{Malygin:1999} is used to determine the change in dislocation density over time:
\begin{equation}
\frac{d n_d}{dt} = \left(\delta_0 \frac{Y_0}{Gb} +  \delta _f \sqrt{n_d} - k_a b n_d \right) |V| n_d
\label{eq:Malyg}
\end{equation}
The first term in parentheses takes into account the generation of dislocations on non-deformation obstacles (impurities, etc.), the second term is the reproduction of dislocations on the ''forest`` dislocations (those crossing the sliding planes) and the third term is the mutual annihilation of dislocations (\cite{Malygin:1999}). This approach has shown itself well in modeling the elastic-plastic behavior of metals under high-pressure loading at strain rates up to $\sim 10^7\un{s^{-1}}$ \cite{Krasnikov:2010}. According to \cite{Estrin:1998} the dislocation density determines the average dislocation cell size $d_c = K/\sqrt{n_d}$, which is inversely related to the square root of the density $n_d$, where $K$ is a constant.

\begin{figure}
\centering
\includegraphics[width=1\columnwidth]{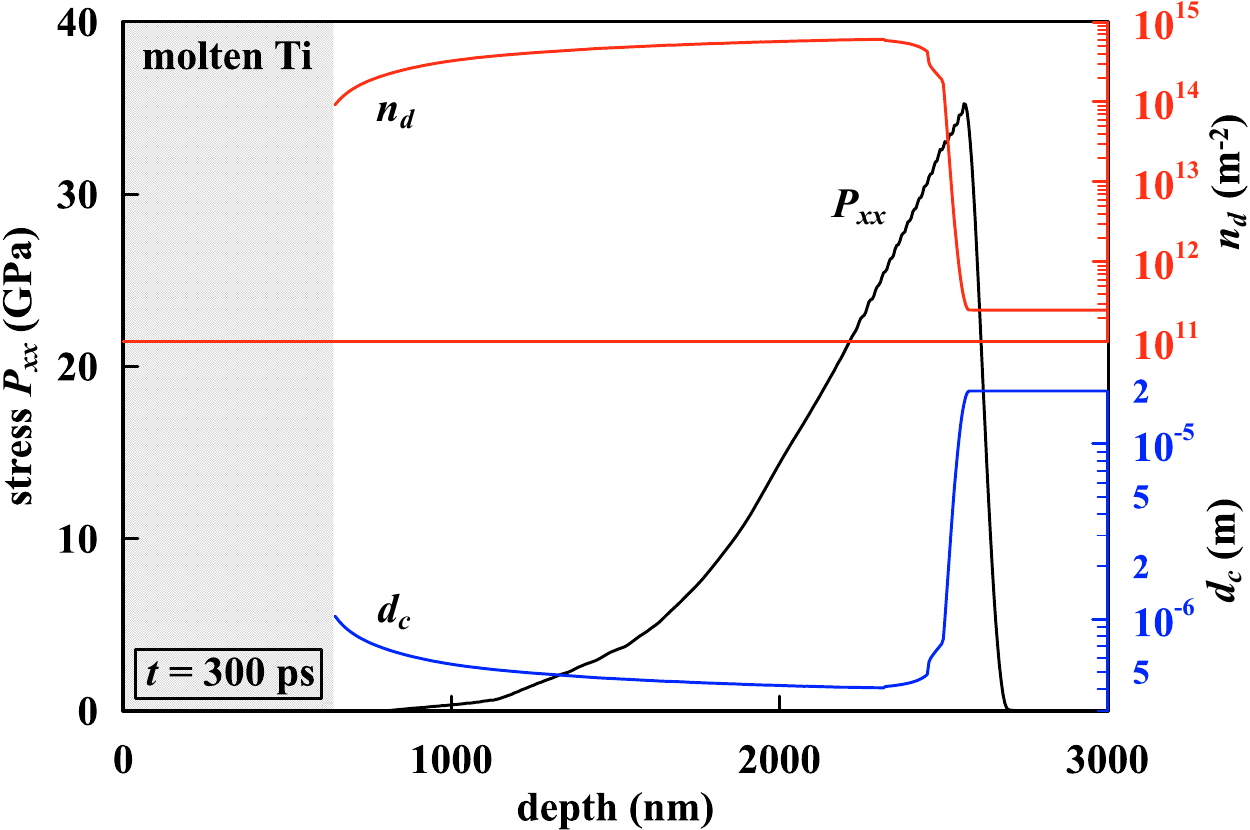}
\caption{\label{fig:dislocation}
Profiles of the longitudinal stress $P_{xx}$, dislocation density $n_d$ and dislocation cell size $d_c$ at the time of 300 ps, when SW transits into the pure elastic regime of propagation resulting in cessation of dislocation production. At this time the depth of modified subsurface layer runs 2500 nm and the dislocation cell size reaches its minimum of 407 nm.
At $T>T_m(p)$ the material melts to a depth of $\sim 640\un{nm}$ from the initial surface. This molten layer is marked by a gray area, where dislocation dynamics is not applicable.
}
\end{figure}

In our model, the initial dislocation density is set equal to the thermodynamically equilibrium density of $10^{11}\un{m^{-2}}$ at the room temperature \cite{Hirth:1968}.
The dislocation-guided yield strength is determined from the Taylor equation $Y = Y_0 + \alpha G b \sqrt{n_d}$ \cite{Suzuki:1991}.
Here the yield strength parameter $Y_0$ is taken to be 10 GPa for consistency with the results of MD simulations presented in the above Sections.

The parameters used in Eq.~(\ref{eq:Malyg}) describing the dislocation dynamics are fitted to the experimental results obtained usually at much lower strain rates than that in our experiment with ultra-short shock loading.
Thus, to reproduce the cell size of $\sim 500 \un{nm}$ obtained in our experimental conditions (see Fig.~\ref{fig:hcp}(c)), it is necessary to use a multiplier of 200 for the model parameters $\delta_0$ and $k_a(V,T)$ proposed in \cite{Krasnikov:2010}.
We believe that the need to increase the $\delta_0$ parameter is due to the formation of titanium particles with $\omega$ phase, which are nucleated during high-rate deformation, as noted in \cite{Nelasov_2021}. Perhaps, such increase of $k_a$ can be also associated with the growth of the concentration of vacancies in a chain of $\alpha-\omega-\alpha$ phase transformations, which was observed experimentally in \cite{Kolobov2016}.

Figure~\ref{fig:dislocation} shows the results of continual modeling with our in-house hydrocode. The profiles are shown at the time of the SW transition from the plastic to elastic regime of deformation, after which the growth of the dislocation density stops. One can see that the maximal depth of subsurface layer with modified titanium is about $2.5\un{\mu m}$ (with respect to the initial sample surface), which agrees well with the thickness $\sim 2\un{\mu m}$ of such layer observed in our experiment.

\section{Conclusion}

We have found with the help of the scanning and transmission electron microscopy that the submicrocrystalline grain-subgrain structure is formed in the thin subsurface layer of technically pure titanium (VT1-0 alloy) been exposed to a powerful femtosecond laser pulse.
The thickness of the modified layer is about $1\un{\mu m}$, and it consists of structure elements elongated in three possible directions: predominantly perpendicular, parallel, and at an angle close to 45 degrees to the surface of irradiated sample. These elements have lengths from $\sim 100\un{nm}$ to several micrometers and their transverse sizes lie in the range $[5 - 200]\un{nm}$. Thus, the modified microstructure is characterized by substantial nonuniformity in the orientation and size distributions of its elements --- there are nanosized (at least in one dimension) grains, as well as the thin grains elongated up to several micrometers along the modified layer. Formation of such grain-subgrain structure suggests that the processes of ultrafast crystallization followed by recrystallization of solid take place within the molten micrometer-sized subsurface layer.

To get insight into the experimental results we have performed two-temperature hydrodynamics modeling and atomistic simulation of cascade of processes induced in titanium after fast heating of electrons in a tiny skin layer. Our simulations indicate that the melting caused by heat propagation involves only top material up to the depth of about 100 nm. The much deeper melting is produced by a strong melting shock wave generated in the heated layer. The initial pressure in such SW may exceed 1 TPa, and it  propagates with the speed about 15 km/s. Large-scale molecular dynamics simulation of single-crystal titanium demonstrates that the SW front continues to melt/liquefy even after its temperature drops below the melting curve $T_m(P)$. The enormous shear stress generated in a narrow SW front leads to collapse/amorphization of the crystal lattice and formation of a supercooled metastable melt. Such melt crystallizes in an unloading tail of SW until its temperature becomes higher than $T_m(P)$ due to a rapid pressure drop. Much later the crystallization of the subsurface molten layer will continue after the heat leaves it. After the shear stress drops below $\sim 12 \un{GPa}$, the cold mechanical melting ceases but the shock-induced plastic deformation continues to modify the solid. The depth of modification is limited by SW attenuation to the Hugoniot elastic limit, and can reach several micrometers.

We have also performed several small-scale MD simulations to study the long-term evolution of the atomic structures obtained after large-scale MD simulation of shock-induced processes. The pressure relaxation and temperature reduction applied to a small shocked sample result in formation of a columnar structure of grains elongated predominately along the direction of maximum heat transfer, which correlates with that observed experimentally.

In contrast to our previous studies of structures formed by exposure to long laser pulses of nanosecond duration, irradiation by a powerful femtosecond pulse leads not only to nanostructuring the thin subsurface layer, but also to formation of cellular deformation structures in a near-surface layer several micrometers in thickness, which exceeds the depth of the melted layer by several times.
Hydrodynamics simulation using a dislocation model also indicates formation of a high-density dislocation layer up to a depth at which the plastic SW front disappears.
Experimentally measured thicknesses of modified subsurface layers produced by a laser-induced melting shock wave are close to their theoretical estimates made in our earlier work \cite{Khokhlov:2022}

The formation of microstructures responsible for significant hardening in the near-surface layers of sufficiently large thickness after femtosecond laser pulse can be used as a method of hardening of bulk metallic components and end products. The entrainment of a thin surface layer, which occurs, for example, in conditions of running-in of machine parts and mechanisms, may be permissible without deterioration of their operational characteristics.

\section*{Acknowledgements}
The experiments on laser exposure were conducted using the Unique Facility “Terawatt Femtosecond Laser Complex” at the Center for Collective Usage “Femtosecond Laser Complex” of the Joint Institute for High Temperatures of the Russian Academy of Sciences (JIHT RAS). This research was supported by The Ministry of Science and Higher Education of the Russian Federation (State Assignment No. 075-01129-23-00, JIHT RAS). Simulations were supported by the State Assignment No. 0029-2019-0003 ''Nonlinear dynamics of complex media`` in L.~D.~Landau ITP RAS.
Analysis of the structural-phase compositions was supported by the FRC PCP and MC RAS in accordance with the state assignment No. AAAA-A19-119111390022-2. TEM images were obtained at
the Center for Collective Usage of FSRC "Crystallography and Photonics" RAS.



%

\end{document}